\documentclass[12pt]{article}
\usepackage{slashed}
\usepackage{epsfig}
\def\be{\begin{equation}}
\def\ee{\end{equation}}
\def\bea{\begin{eqnarray}}
\def\eea{\end{eqnarray}}
\usepackage{graphicx}

\catcode`\@=11
\def\lsim{\mathrel{\mathpalette\@versim<}}
\def\gsim{\mathrel{\mathpalette\@versim>}}
\def\@versim#1#2{\vcenter{\offinterlineskip
\ialign{$\m@th#1\hfil##\hfil$\crcr#2\crcr\sim\crcr } }}
\catcode`\@=12

\catcode`@=12
\begin{document}
\thispagestyle{empty}
\begin{flushright}
UCRHEP-T557\\
November 2015\
\end{flushright}
\vspace{0.6in}
\begin{center}
{\LARGE \bf Type II Radiative Seesaw Model of\\ Neutrino Mass 
with Dark Matter\\}
\vspace{1.2in}
{\bf Sean Fraser, Corey Kownacki, Ernest Ma, and Oleg Popov\\}
\vspace{0.2in}
{\sl Department of Physics and Astronomy, University of California,\\
Riverside, California 92521, USA\\}
\end{center}
\vspace{1.2in}
\begin{abstract}\
We consider a model of neutrino mass with a scalar triplet 
$(\xi^{++},\xi^+,\xi^0)$ assigned lepton number $L=0$, so that the 
tree-level Yukawa coupling $\xi^0 \nu_i \nu_j$ is not allowed. 
It is generated instead through the interaction of $\xi$ and $\nu$ 
with dark matter and the soft breaking of $L$ to $(-1)^L$. 
We discuss the phenomenological implications of this model, including 
$\xi^{++}$ decay and the prognosis of discovering the dark sector 
at the Large Hadron Collider.
\end{abstract}

\newpage
\baselineskip 24pt

\section{Introduction}

Nonzero neutrino mass is necessary to explain the well-established phenomenon 
of neutrino oscillations in many experiments.  Theoretically, neutrino masses 
are usually assumed to be Majorana and come from physics at an energy 
scale higher than that of electroweak symmetry breaking of order 100 GeV.  
As such, the starting point of any theoretical discussion of the 
underlying theory of neutrino mass is the effective dimension-five 
operator~\cite{w79}
\begin{equation}
{\cal L}_5 = - {f_{ij} \over 2 \Lambda} (\nu_i \phi^0 - l_i \phi^+) 
(\nu_j \phi^0 - l_j \phi^+) + H.c.,
\end{equation}
where $(\nu_i,l_i), i=1,2,3$ are the three left-handed lepton doublets of 
the standard model (SM) and $(\phi^+,\phi^0)$ is the one Higgs scalar 
doublet.  As $\phi^0$ acquires a nonzero vacuum expectation value 
$\langle \phi^0 \rangle = v$, the neutrino mass matrix is given by
\begin{equation}
{\cal M}^\nu_{ij} = {f_{ij} v^2 \over \Lambda}.
\end{equation}
Note that ${\cal L}_5$ breaks lepton number $L$ by two units.

It is evident from Eq.~(2) that neutrino mass is seesaw in character, because 
it is inversely proportional to the large effective scale $\Lambda$. 
The three well-known tree-level seesaw realizations~\cite{m98} of 
${\cal L}_5$ may be categorized by the specific heavy particle used to 
obtain it: (I) neutral fermion singlet $N$, (II) scalar triplet 
$(\xi^{++},\xi^+,\xi^0)$, (III) fermion triplet $(\Sigma^+,\Sigma^0,\Sigma^0)$.  
It is also possible to realize ${\cal L}_5$ radiatively in one loop~\cite{m98}
with the particles in the loop belonging to the dark sector, the lightest 
neutral one being the dark matter of the Universe.  The simplest such 
example~\cite{m06} is the well-studied ``scotogenic'' model, from the Greek 
'scotos' meaning darkness.  The one-loop diagram is shown in Fig.~1.
\begin{figure}[htb]
\vspace*{-3cm}
\hspace*{-3cm}
\includegraphics[scale=1.0]{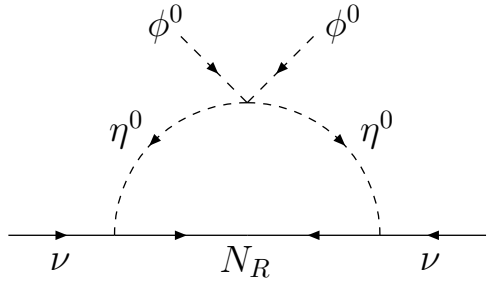}
\vspace*{-21.5cm}
\caption{One-loop $Z_2$ scotogenic neutrino mass.}
\end{figure}
The new particles are a second scalar doublet $(\eta^+,\eta^0)$ and three 
neutral singlet fermions $N_R$.  The dark $Z_2$ is odd for 
$(\eta^+,\eta^0)$ and $N_R$, whereas all SM particles are even. 
This is thus a Type I radiative seesaw model.  It is of course 
possible to replace $N$ with $\Sigma^0$, so it becomes a Type III radiative  
seesaw model~\cite{ms09}.  What then about Type II?

Since ${\cal L}_5$ is a dimension-five operator, any loop realization is 
guaranteed to be finite.  On the other hand, if a Higgs triplet 
$(\xi^{++},\xi^+,\xi^0)$ is added to the SM, a dimension-four coupling 
$\xi^0 \nu_i\nu_j - \xi^+ (\nu_i l_j + l_i \nu_j)/\sqrt{2} + \xi^{++} l_i l_j$ 
is allowed. As $\xi^0$ obtains a small vacuum expectation value~\cite{ms98} 
from its interaction with the SM Higgs doublet, neutrinos acquire small 
Majorana masses, i.e. Type II tree-level seesaw.  
If an exact symmetry is used to forbid this dimension-four coupling, 
it will also forbid any possible loop realization of it.  Hence a Type II 
radiative seesaw is only possible if the symmetry used to forbid the 
hard dimension-four coupling is softly broken in the loop, as 
recently proposed~\cite{m15}.

\section{Type II Radiative Seesaw Neutrino Masses}

The symmetry used to forbid the hard $\xi^0 \nu \nu$ coupling is 
lepton number $U(1)_L$ under which $\xi \sim 0$.   The scalar trilinear 
$\bar{\xi}^0 \phi^0 \phi^0$ term is allowed and induces a small 
$\langle \xi^0 \rangle$, but $\nu$ remains massless.  To connect $\xi^0$ 
to $\nu \nu$ in one loop, we add a new Dirac fermion doublet $(N,E)$ 
with $L=0$, together with three complex neutral scalar singlets $s$ 
with $L=1$.  The resulting 
one-loop diagram is shown in Fig.~2.
\begin{figure}[htb]
\vspace*{-3cm}
\hspace*{-3cm}
\includegraphics[scale=1.0]{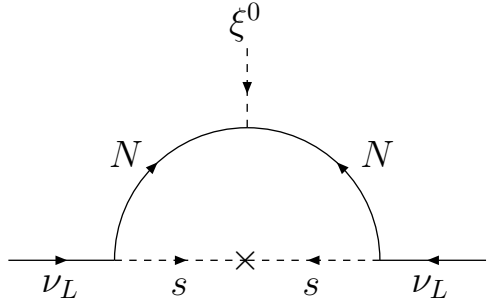}
\vspace*{-21.5cm}
\caption{One-loop neutrino mass from $L=0$ Higgs triplet.}
\end{figure}
Note that the hard terms $\xi^0 N N$ and $s \bar{\nu}_L N_R$ are allowed by 
$L$ conservation, whereas the $s s$ terms break $L$ softly by two units to 
$(-1)^L$.  A dark $Z_2$ parity, i.e. $(-1)^{L+2j}$, exists under which 
$N,E,s$ are odd and $\nu,l,\xi$ are even.  Hence the lightest $s$ is a 
possible dark-matter candidate.  The three $s$ scalars 
are the analogs of the three right-handed sneutrinos in supersymmetry, and 
$(N,E)_{L,R}$ are the analogs of the two higgsinos. However, their 
interactions are simpler here and less constrained.

The usual understanding of the Type II seesaw mechanism is that the scalar 
trilinear term $\mu \xi^\dagger \Phi \Phi$ induces a small vacuum 
expectation value $\langle \xi^0 \rangle = u$ if either $\mu$ is small 
or $m_\xi$ is large or both.  More precisely, consider the scalar potential 
of $\Phi$ and $\xi$.  
\begin{eqnarray}
V &=& m^2 \Phi^\dagger \Phi + M^2 \xi^\dagger \xi + {1 \over 2} \lambda_1 
(\Phi^\dagger \Phi)^2 + {1 \over 2} \lambda_2 (\xi^\dagger \xi)^2 
+ \lambda_3 |2 \xi^{++} \xi^0 - \xi^+ \xi^+|^2 \nonumber \\ 
&+& \lambda_4 (\Phi^\dagger \Phi)(\xi^\dagger \xi) + {1 \over 2} \lambda_5 
[ |\sqrt{2} \xi^{++} \phi^- + \xi^+ \bar{\phi}^0|^2 + 
|\xi^{+} \phi^- + \sqrt{2} \xi^0 \bar{\phi}^0|^2] \nonumber \\ 
&+& \mu (\bar{\xi}^0 \phi^0 \phi^0 + \sqrt{2} \xi^- \phi^0 \phi^+ + 
\xi^{--} \phi^+ \phi^+) + H.c. 
\end{eqnarray}
Let $\langle \phi^0 \rangle = v$, then the conditions for the minimum of $V$ 
are given by~\cite{ms98}
\begin{eqnarray}
m^2 + \lambda_1 v^2 + (\lambda_4 + \lambda_5) u^2 + 2 \mu u &=& 0, \\ 
u[M^2 + \lambda_2 u^2 + (\lambda_4 + \lambda_5) v^2] + 
\mu v^2 &=& 0.
\end{eqnarray}
For $\mu \neq 0$ but small, $u$ is also naturally small because it is 
approximately given by
\begin{equation}
u \simeq {-\mu v^2 \over M^2 + (\lambda_4 + \lambda_5) v^2},
\end{equation}
where $v^2 \simeq -m^2/\lambda_1$.  The physical masses of the $L=0$ Higgs 
triplet are then given by
\begin{eqnarray}
m^2(\xi^0) &\simeq& M^2 + (\lambda_4 + \lambda_5) v^2, \\ 
m^2(\xi^+) &\simeq& M^2 + (\lambda_4 + {1 \over 2}\lambda_5) v^2, \\ 
m^2(\xi^{++}) &\simeq& M^2 + \lambda_4 v^2.
\end{eqnarray}
Since the hard term $\xi^0 \nu \nu$ is forbidden, $u$ by itself does not 
generate a neutrino mass.  Its value does not have to be extremely small 
compared to the electroweak breaking scale.  For example $u \sim 0.1$ GeV 
is acceptable, because its contribution to the precisely measured 
$\rho$ parameter $\rho_0 = 1.00040 \pm 0.00024$~\cite{pdg14} is only 
of order $10^{-6}$.  With the soft breaking of $L$ to $(-1)^L$ shown in Fig.~2,  
Type II radiative seesaw neutrino masses are obtained.  
Let the relevant Yukawa interactions be given by
\begin{equation}
{\cal L}_Y = f_s s \bar{\nu}_L N_R + {1 \over 2} f_R \xi^0 N_R N_R + 
{1 \over 2} f_L \xi^0 N_L N_L + H.c.,
\end{equation}
together with the allowed mass terms $m_E(\bar{N} N + \bar{E} E)$, $m_s^2 
s^*s$, and the $L$ breaking soft term $(1/2)(\Delta m_s^2)s^2 + H.c.$, then 
\begin{equation}
m_\nu = {f_s^2 u r x \over 16 \pi^2} [f_R F_R(x) + f_L F_L(x)],
\end{equation}
where $r = \Delta m_s^2/m_s^2$ and $x=m_s^2/m_E^2$, with
\begin{eqnarray}
F_R(x) &=& {1+x \over (1-x)^2} + {2 x \ln x \over (1-x)^3}, \\ 
F_L(x) &=& {2 \over (1-x)^2} + {(1+x) \ln x \over (1-x)^3}. 
\end{eqnarray}
Using for example $x \sim f_R \sim f_L \sim 0.1$, $r \sim 
f_s \sim 0.01$, we obtain $m_\nu \sim 0.1$ eV for $u \sim 0.1$ GeV.  This 
implies that $\xi$ may be as light as a few 
hundred GeV and be observable, with $\mu \sim 1$ GeV.   
For $f_s \sim 0.01$ and $m_E$ a few hundred GeV, the new contributions 
to the anomalous muon magnetic moment and $\mu \to e \gamma$ are 
negligible in this model.

In the case of three neutrinos, there are of course three $s$ scalars. 
Assuming that the $L$ breaking soft terms $|(\Delta m^2_s)_{ij}| << 
|m^2_{s_i} - m^2_{s_j}|$ for $i \neq j$, then the $3 \times 3$ neutrino mass 
matrix is diagonal to a very good approximation in the basis where the 
$s$ mass-squared matrix is diagonal.  This means that the dark scalars 
$s_j$ couples to $U_{ij} l_i$, where $U_{ij}$ is the neutrino mixing matrix 
linking $e,\mu,\tau$ to the neutrino mass eigenstates $\nu_{1,2,3}$.

\section{Doubly Charged Higgs Production and Decay}

The salient feature of any Type II seesaw model is the doubly charged 
Higgs boson $\xi^{++}$.  If there is a tree-level $\xi^{++} l^-_i l^-_j$ 
coupling, then the dominant decay of $\xi^{++}$ is to $l^+_i l^+_j$. 
Current experimental limits~\cite{atlas14} on the mass of $\xi^{++}$ 
into $e\mu$, $\mu \mu$, and $ee$ final states are about 490 to 550 GeV, 
assuming for each a 100\% branching fraction.  In the present model, 
since the effective $\xi^{++} l^-_i l^-_j$ coupling is one-loop suppressed, 
$\xi^{++} \to W^+ W^+$ should be considered~\cite{kkyy14} instead, for which 
the present limit on $m(\xi^{++})$ is only about 84 GeV~\cite{kkyy15}.  
A dedicated search of the $W^+ W^+$ mode in the future is clearly 
called for.

If $m(\xi^{++}) > 2 m_E$, then the decay channel $\xi^{++} \to E^+ E^+$ 
opens up and will dominate.  In that case, the subsequent decay 
$E^+ \to l^+ s$, i.e. charged lepton plus missing energy, will be the 
signature.  The present experimental limit~\cite{cms14} on $m_E$, 
assuming electroweak pair production, is about 260 GeV if $m_s < 100$ GeV 
for a 100\% branching fraction to $e$ or $\mu$, and no limit if $m_s > 100$ 
GeV.  There is also a lower threshold for $\xi^{++}$ decay, i.e. $m(\xi^{++})$ 
sufficiently greater than $2 m_s$, for which $\xi^{++}$ decays through a 
virtual $E^+ E^+$ pair to $s s l^+ l^+$, resulting in same-sign dileptons plus 
missing energy.

\begin{figure}[htb]
\centering
\includegraphics[scale=.75]{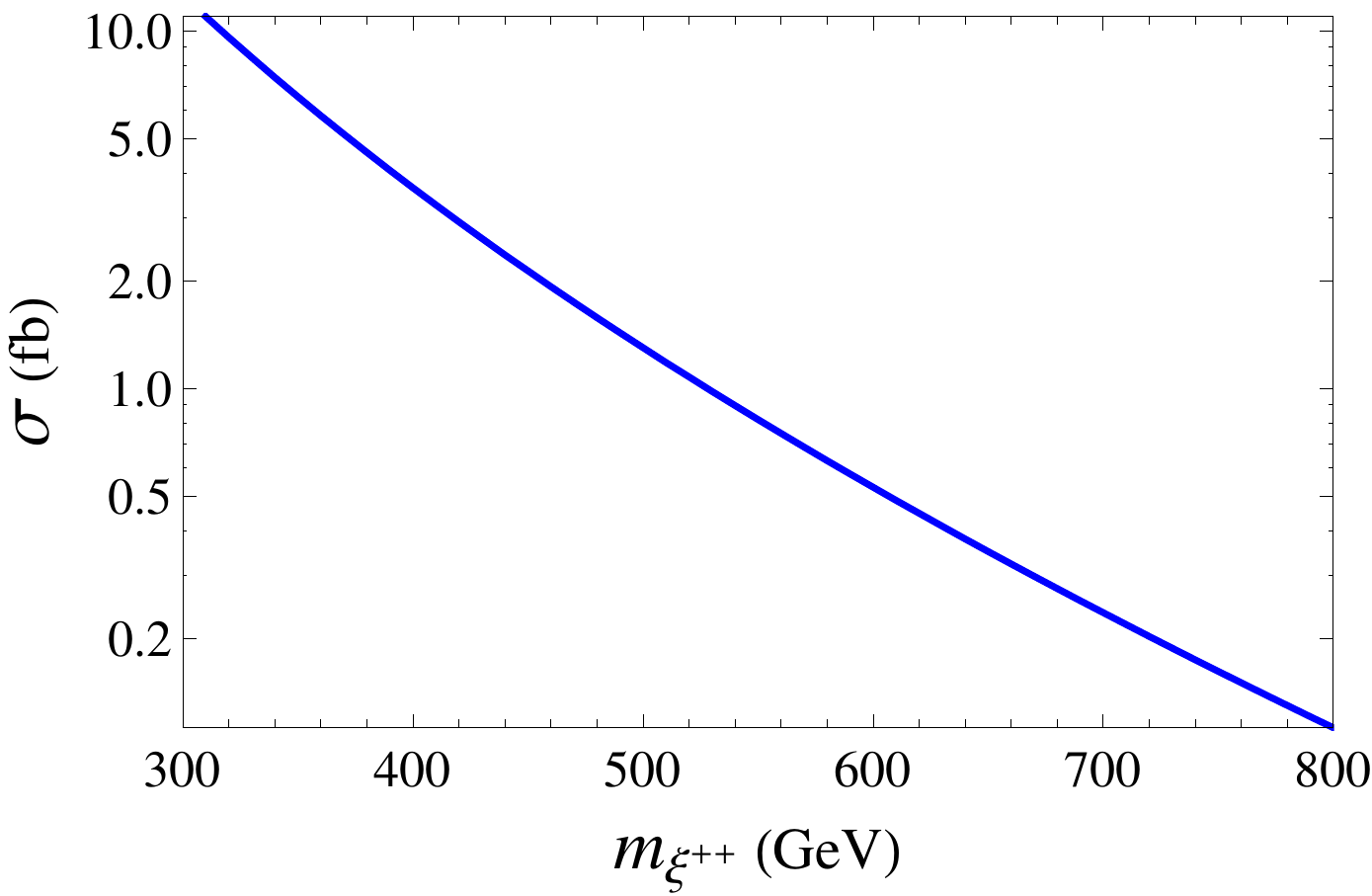}
\caption{LHC Production cross section of $\xi^{++} \xi^{--}$ at 13 TeV.}
\end{figure}

In Fig.~3 we plot the pair production cross section of $\xi^{++} \xi^{--}$ 
at the Large Hadron Collider (LHC) at a center-of-mass energy of 13 TeV. 
We assume that $\xi^+$ and $\xi^0$ are heavier than $\xi^{++}$ so that 
we can focus only on the decay products of $\xi^{\pm \pm}$.  The $W^\pm W^\pm$ 
mode is always possible and should be looked for experimentally in any case. 
However, as already noted, a much more interesting possibility is the case 
$m(\xi^{++}) > 2 m_E$, with the subsequent decay $E^+ \to l^+ s$.  This would 
yield four charged leptons plus missing energy, and depending on the linear 
combination of charged leptons coupling to $s$, there could be exotic final 
states which have very little SM background, becoming thus excellent 
signatures to search for.  Suppose $s_1$ is the lightest scalar, and 
$s_{2,3}$ are heavier than $E^+$, then $E^+$ decays to $s_1 \sum U_{i1} l^+_i$.  
Hence the decay of $\xi^{++} \xi^{--}$ could yield for example $e^+ e^+ 
\mu^- \mu^-$ plus four $s_1$ (missing energy) in the final state.

Recent LHC searches for multilepton signatures at 8 TeV by CMS~\cite{cms} 
and ATLAS~\cite{atlas} are consistent with SM expectations,
and are potential restrictions on our model. 
In particular, the CMS study includes rare SM events
such as $e^+ e^+ \mu^- \mu^-$ and $e^+ e^+ \mu^-$. Due to the absence of
opposite-sign, same-flavor (OSSF) $l^+l^-$ pairs, 
both events are classified as OSSF$0$ where lepton $l$ refers to
electron, muon, or hadronically decaying tau. 
Leptonic tau decays contribute to
the electron and muon counts, and this determines the OSSF$n$ category.
Details from CMS are shown in Table 1
for $\ge 3$ leptons and $N_{\tau{\rm had}}=0$.  
\begin{table}
\begin{center}
\begin{tabular}{|c|c|c|c|c|c|}
\hline 
\multicolumn{6}{|c|}{
Selected CMS results OSSF0 \quad   
$N_{\tau{\rm had}}=0$ , 
$N_b=0$
} \\ 
\hline 
\multicolumn{2}{|c|}{signal regions} & \multicolumn{2}{c|}{$H_T>200$ GeV} & \multicolumn{2}{c|}{$H_T<200$ GeV} \\ 
\hline 
\hline 
$\ge 4$ leptons & $\slashed E_T$ (GeV) & Obs. & Exp.(SM) & Obs. & Exp.(SM) \\  
\hline 
SR1 & $(100,\infty)$ & 0 & $0.01 ^{+0.03}_{-0.01}$ & 0 & $0.11 ^{+0.08}_{-0.08}$\\ 
\hline 
SR2 & $(50,100)$ & 0 & $0.00 ^{+0.02}_{-0.00}$ & 0 & $0.01 ^{+0.03}_{-0.01}$\\ 
\hline 
SR3 & $(0,50)$ & 0 & $0.00 ^{+0.02}_{-0.00}$  & 0 & $0.01 ^{+0.02}_{-0.01}$\\ 
\hline
\hline
3 leptons & $\slashed E_T$ (GeV) & Obs. & Exp.(SM) & Obs. & Exp.(SM) \\ 
\hline
SR4 & $(100,\infty)$ & 5 & $3.7 \pm 1.6$ & 7 & $11.0 \pm 4.9$\\ 
\hline 
SR5 & $(50,100)$ & 3 & $3.5 \pm 1.4$  & 35 & $38 \pm 15$\\ 
\hline 
SR6 & $(0,50)$ & 4 & $2.1 \pm 0.8$  & 53 & $51 \pm 11$\\ 
\hline
\end{tabular} 
\caption{Events observed by CMS at 8 TeV with integrated luminosity 
$19.5$ fb$^{-1}$.} 
\end{center}
\end{table}
The CMS study estimates a negligible SM background for 
SR1-SR3, and in our simulation we use the same selection criteria.
We impose the cuts on transverse momentum
$p_T >$~10 GeV and psuedorapidity
$|\eta| < 2.4$
for each charged lepton, 
with at least one lepton $p_T >$~20 GeV.
In order to be isolated, each lepton with $p_T$ must satisfy 
$\sum_i p_{Ti}<0.15 p_T$, where the sum 
is over all objects within 
a cone of radius $\Delta R = 0.3$
around the lepton direction.

We implement our model with 
FeynRules~2.0~\cite{fr}. 
Using the CTEQ6L1 parton distribution functions,  
we generate events using 
MadGraph5~\cite{mg5},
which includes
the Pythia package for hadronization and showering.  
MadAnalysis~\cite{ma5} is then used with the Delphes card designed 
for CMS detector simulation. 
Generated events intially have 4 leptons. About half
are detected as 3 lepton events,
but the constraints from 
signal regions SR4-SR6 are less restrictive
than SR1-SR3.
The number of detected events in  
the OSSF0 $\ge$ 4 lepton category
is almost the same as
$e^{\pm} e^{\pm} \mu^{\mp} \mu^{\mp} 2 s_1 2 s_1^* $
with very few additional leptons from showering or 
initial/final state radiation.

To examine the production of $e^{\pm} e^{\pm} \mu^{\mp} \mu^{\mp}$ 
we take the mass of $s_1$ to be 130 GeV, which allows $s_1$ to be dark 
matter as discussed in the next section.  
We use the values $f_R = f_L = 0.1$ and $f_s = 0.01$, 
although the results are not sensitive to the exact values 
due to on-shell production and decay.
The effects due to $u \sim 0.1$ GeV may be neglected.  

For our model, we scan the mass range of $\xi^{++}$ and $E^+$. 
In Fig.~4 we plot contours showing the expected number of 
detected events in  
the OSSF0 $\ge$ 4 lepton category
for 13 TeV at luminosity 100 fb$^{-1}$ assuming 
a negligible background as for the 8 TeV case. 
Although the branching fractions of $E^+$ to
$\tau^+ s_1$ or $\mu^+ s_1$ are comparable, 
we find that most of the contributions from $\tau^{\pm}$ decay 
to $e^{\pm}$ or $\mu^{\pm}$ in the $\ge$ 4 lepton final state are not detected.
A similar analysis performed for 8 TeV at 19.5~fb$^{-1}$ 
has a maximum number of detected events of 0.4 
in the plot analogous to Fig. 4, which corresponds to a small
estimated exclusion at the 15\% confidence level.

\begin{figure}[htb]
\centering
\includegraphics[scale=.8]{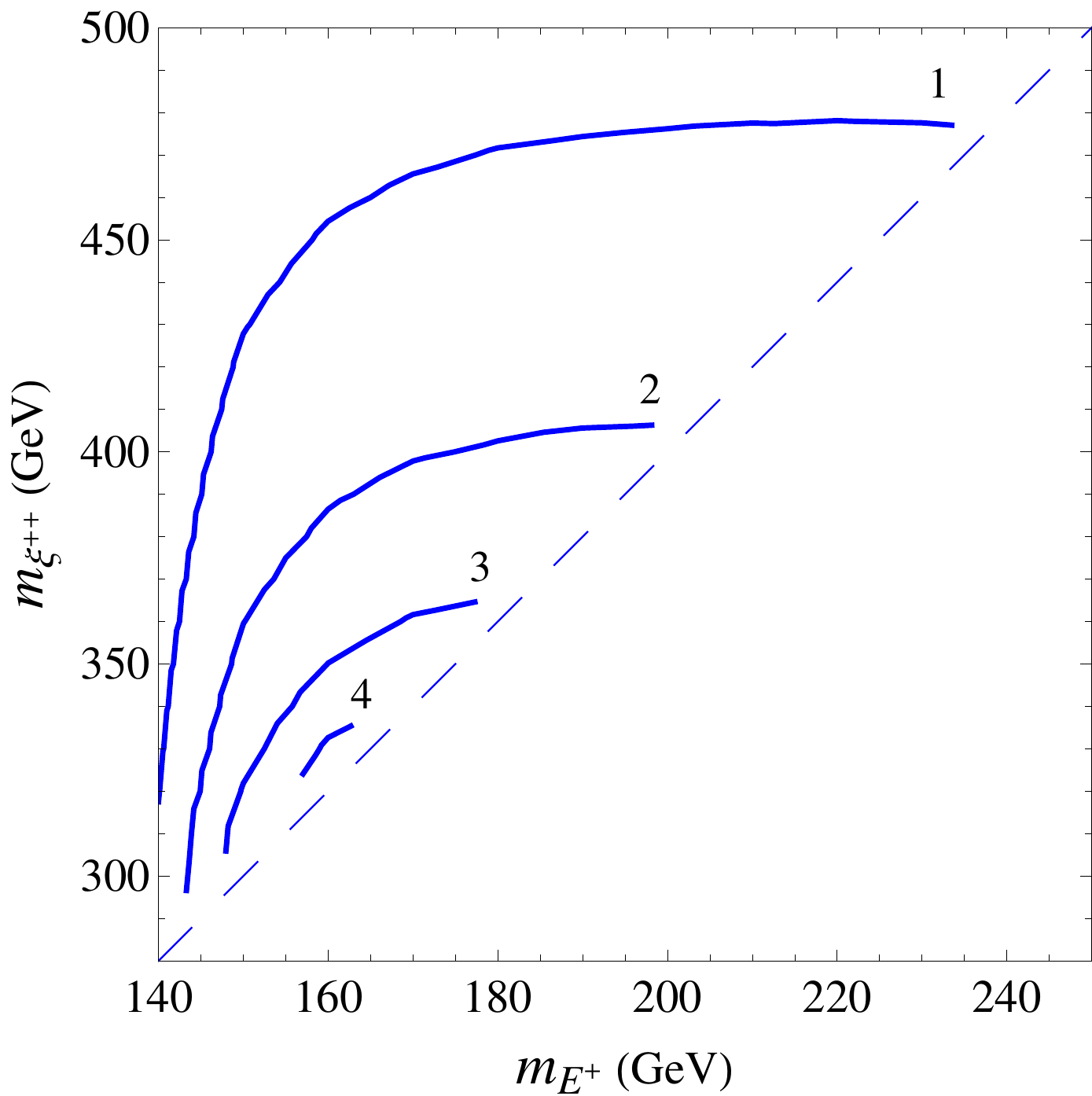}
\caption{
Number of $e^{\pm} e^{\pm} \mu^{\mp} \mu^{\mp} 2 s_1 2 s_1^*$
events for 13 TeV at luminosity 100 fb$^{-1}$.
}
\end{figure}

\section{Dark Matter Properties}

The lightest $s$, say $s_1$, is dark matter.  Its interaction with leptons 
is too weak to provide a large enough annihilation cross section to 
explain the present dark matter relic density $\Omega_M$ of the Universe.  
However, it also interacts with the SM Higgs boson through the usual quartic 
coupling $\lambda_s s^* s \Phi^\dagger \Phi$.  For a value of $\lambda_s$ 
consistent with $\Omega_M$, the direct-detection cross section in 
underground experiments is determined as a function of $m_s$.  A recent 
analysis~\cite{fpu15} for a real $s$ claims that the resulting allowed 
parameter space is limited to a small region near $m_s < m_h/2$.

In our model, we can evade this constraint by evoking $s_{2,3}$.  The 
mass-squared matrix spanning $s_i^* s_j$ is given by
\begin{equation}
({\cal M}^2_s)_{ij} = m^2_{ij} + \lambda_{ij} v^2,
\end{equation}
whereas the coupling matrix of the one Higgs $h$ to $s_i^* s_j$ is
$\lambda_{ij} v \sqrt{2}$.  Upon diagonalizing ${\cal M}^2_s$, the 
coupling matrix will not be diagonal in general.  In the physical basis, 
$s_1$ will interact with $s_2$ through $h$.  This allows the annihilation 
of $s_1 s_1^*$ to $hh$ through $s_2$ exchange, and contributes to $\Omega_M$ 
without affecting the $s_1$ scattering cross section off nuclei through $h$. 
This mechanism restores $s_1$ as a dark-matter candidate for $m_s > m_h$.
\begin{figure}[htb]
\centering
\includegraphics[scale=.8]{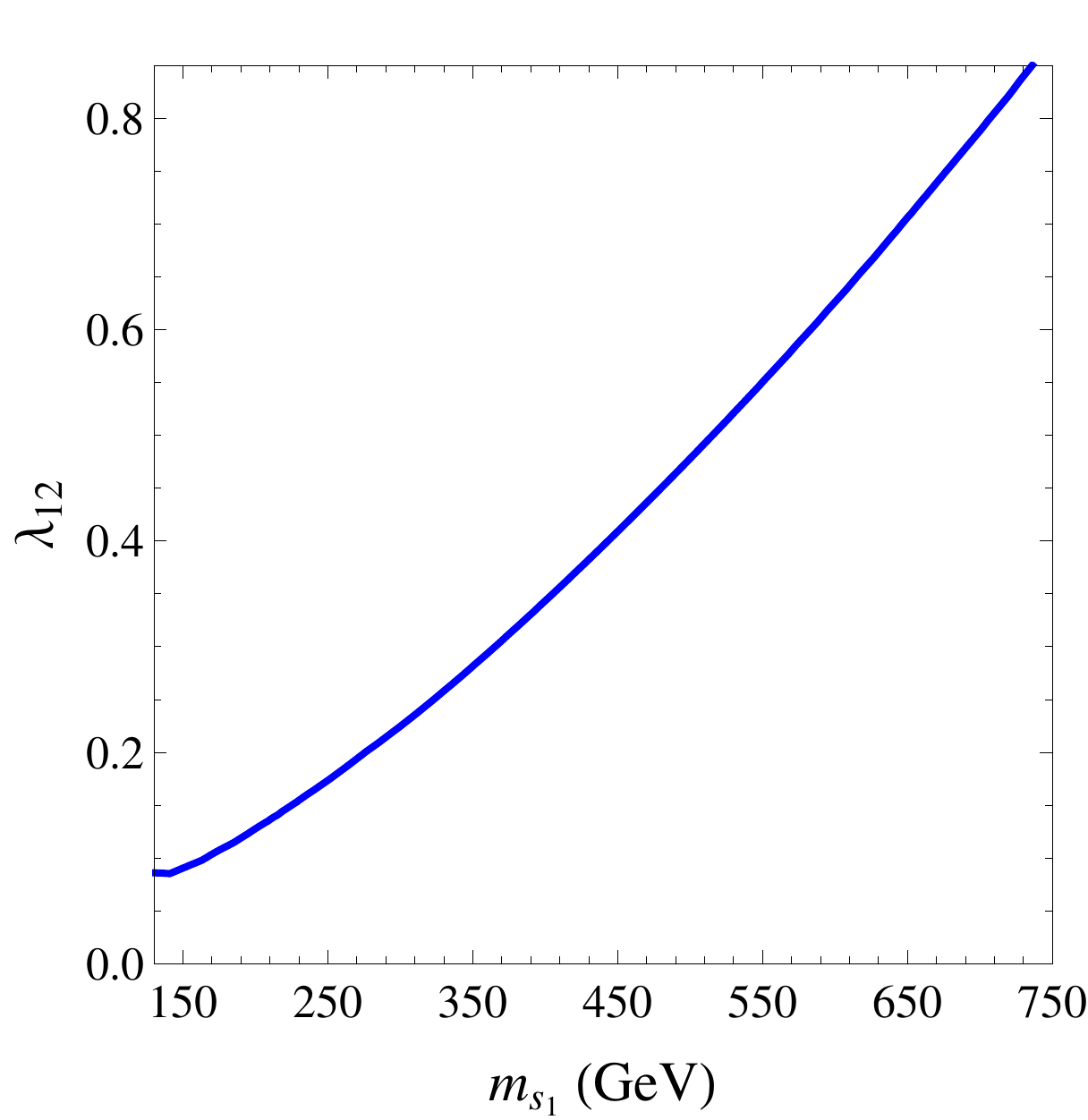}
\caption{
Allowed values of 
$\lambda_{12}$
plotted against $m_{s_1}$
from relic abundance assuming $\lambda_{11}=0$.
}
\end{figure}
 
To demonstrate the scale of the values involved, 
we consider the simplifying case when 
$m_{s_2}=m_{s_3}$ and $\lambda_{12}=\lambda_{13}$.
The additional choice 
$m_{s_{2,3}}^2=m_{s_1}^2+m_h^2$
ensures that $s_{2,3}$ are heavier than $s_1$, and is convenient
because then the relic abundance requirement no longer depends 
explicitly on $m_{s_{2,3}}^2$.
Taking into account that $s_1$ is a complex scalar, we use 
$\sigma  \times v_{rel}$~=~4.4~$\times 10^{-26}$cm$^3$s$^{-1}$~\cite{relic}
and in Fig.~5 we plot the allowed values for 
$\lambda_{12}$ and $m_{s_1}$ taking $\lambda_{11}=0$ for
simplicity to satisfy the LUX data. 

Another possible scenario is to add a light scalar $\chi$ with $L=0$, 
which acts as a mediator for $s$ self-interactions.  This has important 
astrophysical implications~\cite{ss00,m02,m04,k12,kly15,t14}. In this case, 
$s_1 s_1^*$ annihilating to $\chi \chi$ becomes possible.

\section{Conclusion}

We have studied a new radiative Type II seesaw model of neutrino mass with 
dark matter~\cite{m15}, which predicts a doubly charged Higgs boson $\xi^{++}$ 
with suppressed decay to $l^+ l^+$, thereby evading the present LHC bounds 
of 490 to 550 GeV on its mass.  In this model, $\xi^{++}$ may decay to 
two charged heavy fermions $E^+ E^+$, each with odd dark parity.  The 
subsequent decay of $E^+$ is into a charged lepton $l^+$ and a scalar $s$ 
which is dark matter.  Hence there is the interesting possibility of 
four charged leptons, such as $\mu^- \mu^- e^+ e^+$, plus large missing 
energy in the final state.  We show that the LHC at 13 TeV will be able 
to probe such a doubly charged Higgs boson with a mass of the order 
400 to 500 GeV.

\medskip
This work is supported in part 
by the U.~S.~Department of Energy under Grant No.~DE-SC0008541.

\baselineskip 18pt
\bibliographystyle{unsrt}

\end{document}